\documentclass[aps,prb,twocolumn,nofootinbib,floatfix,bibliography,superscriptaddress]{revtex4-2}
\usepackage{graphics,bm,amsmath,amssymb,natbib,url,epsfig,psfrag,color}
\usepackage{graphicx}
\graphicspath{{./Figs/}{./}} 
\usepackage{braket}
\usepackage[caption=false]{subfig}
\newcommand{\phantomsubfloat}[1]{
    {
        \captionsetup[subfigure]{labelformat=empty}
        \subfloat[][]{#1}     }%
} 

\newcommand{\be}{\begin{equation}}
\newcommand{\ee}{\end{equation}}
\newcommand{\bea}{\begin{eqnarray}}
\newcommand{\eea}{\end{eqnarray}}

\newcommand{\beq}{\begin{eqnarray}}
\newcommand{\eeq}{\end{eqnarray}}
\usepackage{xcolor}
\usepackage[unicode=true,pdfusetitle,bookmarks=true,bookmarksnumbered=false,bookmarksopen=false,breaklinks=false,pdfborder={0 0 0},pdfborderstyle={},backref=false,colorlinks=true]{hyperref}
\hypersetup{colorlinks,allcolors=blue}

\colorlet{mlc}{orange!70!gray}
\newcommand{\MLc}[1]{\textcolor{mlc}{[ML: {\it #1}\,]}}

\newcommand{\ZM}[1]{{\color{red}#1}}
\begin{document}
\newcommand{\Tr}{\text{Tr}}
\newcommand{\eran}[1]{{\color{blue}#1}}

\def\bs#1\es{\begin{split}#1\end{split}}	\def\bal#1\eal{\begin{align}#1\end{align}}
\newcommand{\nn}{\nonumber}
\newcommand{\sgn}{\text{sgn}}

\title{Realizing the interacting resonant level model using a quantum dot detector}
	
\author{Zhanyu Ma}
\affiliation{Raymond and Beverly Sackler School of Physics and Astronomy, Tel Aviv University, Tel Aviv 69978, Israel}

\author{Matan Lotem}
\affiliation{Department of Physics, Ben-Gurion University of the Negev, Beer-Sheva, 84105 Israel}

\author{Yigal Meir}
\affiliation{Department of Physics, Ben-Gurion University of the Negev, Beer-Sheva, 84105 Israel}

\author{Eran Sela}
\affiliation{Raymond and Beverly Sackler School of Physics and Astronomy, Tel Aviv University, Tel Aviv 69978, Israel}

\date{\today}
\begin{abstract}
The interacting resonant level model (IRLM) is the simplest quantum impurity model to display strongly correlated effects in mesoscopic systems, which triggered its extensive theoretical study. However, to date, there have not been any realizations of the model with  controllable interaction parameter, and thus the detailed predictions could not be confirmed. Here we use a recently developed approach to Anderson orthogonality catastrophe physics, using a charge detector coupled to a quantum dot (QD) system, to devise a simple experimental system which could display IRLM behavior, and detail its predictions. At the same time, the mapping to IRLM allows to determine the interaction parameter of the charge detector using simple experimental probes.
\end{abstract}
\maketitle

\section{Introduction} 
 The interacting resonant level model (IRLM) is given crudely by the Hamiltonian 
 \be
 \label{eq:IRLM}
 \delta H=t'(\psi^\dagger d + \mathrm{H.c.})+ U \psi^\dagger \psi d^\dagger d,
 \ee
and describes a resonant $d$-level coupled both by tunneling and by interactions to its fermionic bath. The effects of the interaction on physical properties encode nontrivial renormalization group physics. Particularly, the resonance width is a key energy scale denoted $T_K$, which scales as a power-law of the tunneling amplitude, $T_K \sim {t'}^\alpha$. This energy scale controls the behavior of various physical quantities. For example, the charge susceptibility is inversely proportional to $T_K$ and the thermodynamic entropy drops with decreasing temperature from $\ln(2)$ to 0 at $T \sim T_K$.

The IRLM appeared extensively in the literature as a testbed for renormalization group methods and exact solutions both at equilibrium and out of equilibrium~\cite{PhysRevLett.96.216802,PhysRevB.75.125107,PhysRevB.78.201301,PhysRevB.77.033409,PhysRevLett.101.140601,PhysRevLett.105.146805,PhysRevLett.107.206801,PhysRevB.90.155110,PhysRevB.89.081401, PhysRevB.87.195112, PhysRevB.87.075319}. Being related to elementary quantum impurity models such as the anisotropic Kondo model~\cite{finkelstein,PhysRevB.25.4815}, the spin-boson model~\cite{guinea1985bosonization,PhysRevB.93.165130}, see also~\cite{kashcheyevs2009quantum,goldstein2009interacting,rylands2017quantum, PhysRevB.96.075435}, it captures the basic physics of Anderson orthogonality catastrophe and Fermi edge singularity. Thus, it is fair to say that the IRLM featured in various experiments, see for example~\cite{Mebrahtu_2012}. Yet, while Eq.~(\ref{eq:IRLM}) is a generic model of a localized site to a metallic medium, in practice it is not easy to tune $t'$ and $U$ separately and at large coupling $t'$  the interaction is exceedingly screened. Thus, a controllable realization of the IRLM remains desirable.

Here we revisit our recent work on a double quantum dot coupled to a charge detector, which is a widely studied system in mesoscopic physics \cite{ihn2009semiconductor}, serving as spin \cite{Harvey_2022} or charge \cite{gorman2005charge} qubits and giving a paradigmatic model to study dephasing \cite{PhysRevB.56.15215} and backaction effects of the detector \cite{kung2009noise,Bischoff_2015,biesinger2015intrinsic,ferguson2023measurement} either in- or out of equilibrium. We argued that
a localization transition~\cite{Ma_2023} can be observed in such a setup at equilibrium when the electrostatic coupling to the detector is large enough. Interestingly, we will show that the model studied in~\cite{Ma_2023} maps to the IRLM, thus relating the pair of parameters $t'$ and $U$ to the various model parameters of the double dot coupled electrostatically to the detector. Therefore the physics of the energy scale $T_K$ and its nontrivial scaling can be observed in a double quantum dot coupled to a charge detector.

Our motivation for this relationship is to identify experimental observables that can help to determine the strength of the interaction between the detector and the double dot: (i) entropy measurements - which became possible due to recent experimental progress \cite{Hartman_2018,Child_2022}, (ii) charge measurements -- which can reveal the charge susceptibility of quantum impurity problems \cite{Piquard_2023}, or (iii) transport measurements. The scaling of $T_K$ with tunneling \cite{Camacho_2019} in all of these quantities allows them to determine the relevant interaction strength.  We note that there are additional ways to gauge the strength of the electrostatic coupling. In  Ref.~\cite{sankar2024detector} we considered experiments such as~\cite{kung2009noise,ferguson2023measurement} in the weak-tunneling regime. Here we focus on the effects of the detector in the strong tunneling regime where coherence effects develop below a certain scale $T_K$.


The paper is organized as follows. The triple QD model studied throughout this work, which consists of a double dot electrostatically coupled to a detector QD, is introduced in Sec.~\ref{se:model} and mapped to the IRLM in the limit of large electrostatic coupling using a Schrieffer-Wolff transformation. In Sec.~\ref{se:Tk} we extend the mapping to the IRLM into a broader parameter range using numerical renormalization group (NRG) and and extract the emergent energy scale $T_K$ from the thermodynamic entropy. While it is possible to numerically extract $t'$ and $U$, thus establishing the mapping, from the NRG level flow of the original model\cite{hewson2004renormalized, PhysRevB.87.195112, PhysRevB.87.075319, PhysRevB.90.155110}, to avoid the slightly cumbersome calculations, here we instead focus on the universal low energy properties such as $\alpha$, which can be easily read out from the scaling relation $T_K\sim t'^{\alpha}$. We then proceed to discuss physical quantities: in Sec.~\ref{se:charge-susceptibility} we discuss the charge susceptibility and in Sec.~\ref{se:spectral_function} we study the conductance. We conclude with open questions in Sec.~\ref{se:summary}.

\begin{figure*}
\centering
\includegraphics[width=1\textwidth]{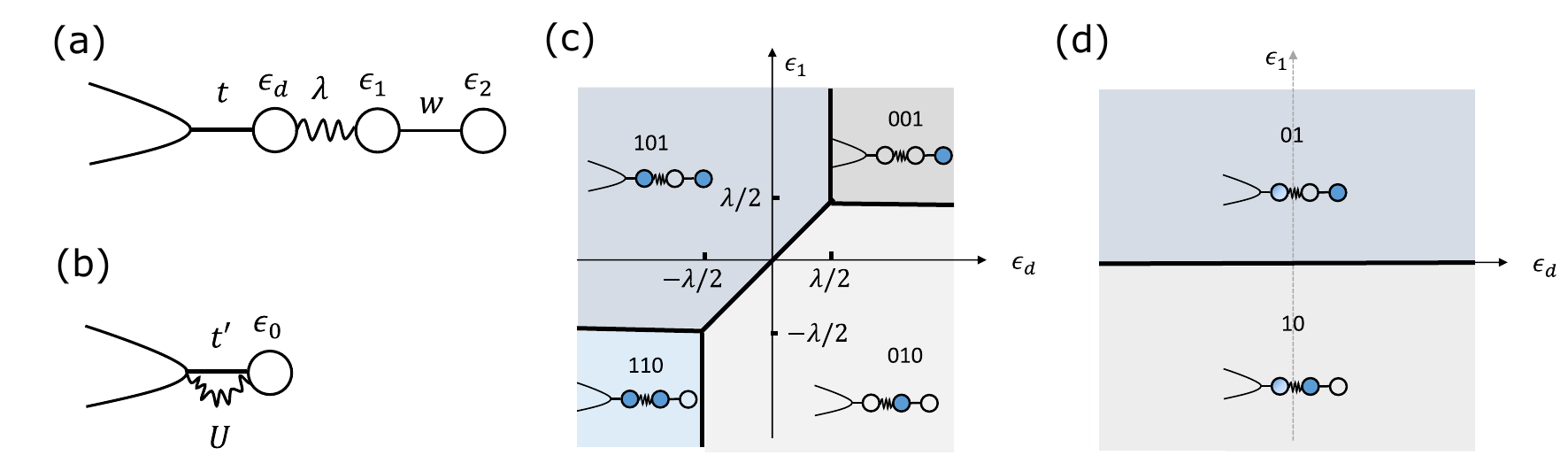}
  \vspace{-2\baselineskip}\phantomsubfloat{\label{fg:Model}}\phantomsubfloat{\label{fg:IRLM}}\phantomsubfloat{\label{fg:Charge-L}}\phantomsubfloat{\label{fg:Charge-G}}
  \caption{(a) Quantum dot (QD) detector coupled to a double dot consisting of QD1 and QD2.  (b) Interacting resonant level model (IRLM). (c) Charge-stability diagram at $t=w=\epsilon_2=0$ (and in particular $\lambda\gg\Gamma$). The three digits denote the occupation of the $d$-level, QD1, and QD2, i.e., $\ket{n_d,n_1,n_2}$, and we assume that a single electron resides in the double dot. (d) Charge-stability diagram for large $\Gamma$, so that the charge state of the $d$-level is not well defined, and the states are denoted by the charge of QD1 and QD2, i.e., $\ket{n_1,n_2}$.}
    \label{fg:1}
\end{figure*}


\section{Model and mapping to IRLM}
\label{se:model}
We consider a spinless model of a double dot electrostatically coupled to a QD detector, as shown in Fig.~\ref{fg:Model}:
\bea\label{eq:H}
    H &=& H_\text{lead}[\psi] 
       +  \overbrace{t[\psi^\dagger(0)d+\mathrm{H.c.}]}^{H_t} 
       +  \epsilon_d d^\dagger d \notag \\
      &+& \lambda(d^\dagger d-\tfrac{1}{2})(d_1^\dagger d_1^{\,}-\tfrac{1}{2}) \\
      &+&  \underbrace{w[d_1^\dagger d_2^{\,}+\mathrm{H.c.}]}_{H_w}
       +  \sum_{j=1,2}\epsilon_j d_j^\dagger d_j^{\,} . \notag
\eea
The first line describes the QD detector, consisting of the lead fermions $\psi$ and the $d$-level operator $d$. The lead Hamiltonian is written as a chiral fermion
\be \label{eq:H-lead}
    H_\text{lead}[\psi] = -i\hbar v_F\int_{-\infty}^\infty dx\psi^\dagger\partial_x\psi,
\ee
with $\{ \psi(x),\psi^\dagger(x') \}=\delta(x-x')$. We define $\Gamma$ as the width of the $d$-level in the absence of the subsequent lines of $H$ and express it in terms of the lead density of states $\nu$, 
\be
    \Gamma\equiv\pi\nu t^2,\ \ \ \nu=\frac{1}{2\pi\hbar v_F}.
\ee
The double dot consists of fermion operators $d_1$ and $d_2$, for QD1 and QD2, respectively, that are tunnel coupled with amplitude $w$. The left dot is electrostatically coupled to the QD detector with strength $\lambda$. Throughout we assume a single electron resides in the double dot, and so it is sufficient to set $\epsilon_2=0$.

In the limit of $w=0$ we have a  resonant level model (RLM) with $\lambda$ entering as an  energy shift of the $d$-level, that is $\epsilon_d\to\epsilon_d \pm \lambda/2$ for $d^\dagger_1 d_1^{\,}=1$ or $0$. Following Ref.~\cite{Ma_2023}, we define $\delta_\text{DD}$ as half the phase shift difference between these two states,
\begin{subequations}
\bea\label{eq:delta-DD}
    \delta_\text{DD} =&& \frac{1}{2}\Big[\delta\big(\epsilon_d{+}\tfrac{\lambda}{2},\Gamma\big){-}\delta\big(\epsilon_d{-}\tfrac{\lambda}{2},\Gamma\big)\Big], \\
    &&\delta(\epsilon_d,\Gamma) = \arctan\frac{\epsilon_d}{\Gamma}.\label{eq:delta-RLM}
\eea
\end{subequations}
 For the QD detector setup, as the interaction $\lambda$ goes from weak to strong, the phase shift $\delta_\text{DD}$ goes from $0$ to $\pi/2$. Thus, it cannot reach the critical value $\delta_c=\pi/\sqrt 2$ of the localization transition discussed in Ref.~\cite{Ma_2023}.

Re-introducing $w$, and excluding specific finely-tuned degeneracy points, the low-energy physics of the double-dot model can be described as a two-level system coupled to a dissipative bath. A variety of equivalent models can be used to describe such systems (see e.g., Ref.~\cite{PhysRevB.93.165130}). Here we find it insightful to represent the system as an IRLM, as depicted in Fig.~\ref{fg:IRLM}
\bea\label{eq:H-IRLM}
    H_\text{IRLM}
      &=& H_\text{lead}[\psi'] 
       + t'[\psi'^\dagger(0)\tilde{d} + \mathrm{H.c.}] 
       + \epsilon_0 \tilde{d}^\dagger\tilde{d} \notag\\
    &+& U(\psi'^\dagger(0)\psi'(0)-\tfrac{1}{2})
         (\tilde{d}^\dagger\tilde{d}-\tfrac{1}{2}).
\eea
The fermionic lead is given as in Eq.~\eqref{eq:H-lead}, and we denote the fermionic field  $\psi'$  and level $\tilde{d}$, to stress that they could differ from the $\psi$ and $d$ of the original model in Eq.~\eqref{eq:H}.  

We can equivalently parameterize the interaction of the IRLM by $U$ or by the phase shift
\be\label{eq:delta-IRLM}
    \delta = \arctan\frac{\pi\nu U}{2}.
\ee
The latter is more convenient since the precise meaning of $U$ 
depends on cut-off conventions [see Eqs.~(25.29) and (25.34) in Ref.~\cite{gogolin2004bosonization}].

As will be shown in the following subsections, the relation between the IRLM phase shift and $\delta_\text{DD}$ of the double-dot model is given by
\be\label{eq:delta}
    \delta  = \frac{\pi}{2}-\delta_\text{DD} \overset{(\epsilon_d=0)}{=} \arctan\frac{\Gamma}{\lambda/2}.
\ee
Observe that weak interaction in the double-dot model corresponds to strong interaction of the IRLM and vice-versa, which might lead to confusion. Indeed, in the language of the original double-dot model, the zero-interaction limit $\lambda \ll \Gamma$ yields a finite phase shift $\delta=\pi/2$. But this follows because  $\delta=0$ corresponds to a decoupled lead ($\Gamma\to 0$), while finite $\Gamma$ with $\lambda\to 0$ means that we have extended the lead to include one more site. In the IRLM picture, this definition of the phase shift is natural, with $\delta=0$ corresponding to the zero-interaction $U=0$ limit and  $\delta=\pi/2$ corresponding to the $U\to \infty$ limit.
Note that in contrast to the double-dot model with a QD detector, the IRLM does undergo a Kosterlitz-Thouless transition. However, this occurs at negative $U$ (or $\delta$ -- see e.g., \cite{PhysRevB.93.165130}), whereas in our case $\delta$ is always positive. A Kosterlitz-Thouless transition at positive $U$ is possible for more than 3 leads \cite{Camacho_2019}.

While the mapping of the double-dot model to the IRLM with $\delta$  as in Eq.~\eqref{eq:delta} holds generally, here we will only derive it in the two limits of $\lambda\gg\Gamma$ (Sec.~\ref{se:SW1}) and $\lambda\ll\Gamma$ (Sec.~\ref{se:SW2}). We will then verify this relation numerically for the full range of $\lambda/\Gamma$, and discuss its consequences in Sec.~\ref{se:Tk}.

\subsection{Mapping to the IRLM: $\lambda \gg \Gamma$}\label{se:SW1}
In this limit, we start from the reference point of $w=\Gamma=0$. Then we can work with well-defined occupation states for the $d$-level and the two dots, that we denote by  $\ket{n_d n_1 n_2}$ with $n_{j=d,1,2}=0,1$ indicating empty or occupied. Fixing the double dot to single occupation and setting $\epsilon_2=0$, we have four states with energies
\bea
    E_{101}&=&-\frac{\lambda}{4}+\epsilon_d,~~~~~~~~~
    E_{010}=-\frac{\lambda}{4}+\epsilon_1, \nonumber \\
    E_{110}&=&+\frac{\lambda}{4}+\epsilon_d+\epsilon_1, ~~~
    E_{001}=+\frac{\lambda}{4}.
\eea
These are depicted in the charge stability diagram in Fig.~\ref{fg:Charge-L}, indicating which of the four is the ground state as a function of $\epsilon_d$  and $\epsilon_1$. Along the diagonal transition line, we have two degenerate states, $\ket{101}$ and $\ket{010}$. As we will now show, when coupled to the lead, these form the two states of a $\tilde{d}$-level in the IRLM of Eq.~\eqref{eq:H-IRLM}. At the origin of the charge-stability diagram, we have particle-hole symmetry. Throughout we will focus on this point, unless explicitly stated otherwise (e.g., as in Sec.~\ref{se:charge-susceptibility}), setting $\epsilon_d=\epsilon_1=\epsilon_2=0$.

We treat all the tunneling terms $H_t$ and $H_w$ perturbatively, defining $H_0 \equiv H-H_w-H_t$. Performing a Schrieffer-Wolff \cite{SchriefferWolff} transformation with respect to $w$ and $t$, yields an effective Hamiltonian
\be
    \delta H =(H_t+H_w)\frac{1}{E-H_0}(H_t+H_w),
\ee
where $E=-\lambda/4$ is the ground state energy to zeroth order in $t,w$ and we implicitly assume a projection onto the ground space. This yields 3 types of terms,
\bea
    \delta H_{t^2}&=& -\frac{t^2}{\lambda/2}
        \Big(\ket{101}\bra{101}d^\dagger\psi\psi^\dagger d\ket{101}\bra{101} \notag \\
        &&~~~~~~~+\ket{010}\bra{010}\psi^\dagger d d^\dagger\psi\ket{010}\bra{010}\Big) \notag \\
        &=& -\frac{t^2}{\lambda/2}
            \Big(\psi\psi^\dagger\underbrace{\ket{101}\bra{101}}_{\tilde{d}^\dagger\tilde{d}}
            + \psi^\dagger\psi \underbrace{\ket{010}\bra{010}}_{\tilde{d}\tilde{d}^\dagger} \Big), \notag \\
    \delta H_{w^2}&=&-\frac{w^2}{\lambda/2} = \text{const}, \\
    \delta H_{tw} &=& -\frac{tw}{\lambda/2} 
    \Big(\ket{010}\bra{010} \psi^\dagger d d^\dagger_1 d_2 \ket{101}\bra{101} \notag \\
        &&~~~~~~~+\ket{101}\bra{101}d^\dagger\psi d^\dagger_2 d_1 \ket{010}\bra{010} \Big) + \mathrm{H.c.}\notag \\
        &=&-\frac{tw}{\lambda/2} \Big(\psi^\dagger \underbrace{\ket{010}\bra{101}}_{\tilde{d}} + \underbrace{\ket{101} \bra{010}}_{\tilde{d}^\dagger}\psi\Big) + \mathrm{H.c.} \notag
\eea
Thus, we obtained the IRLM of Eq.~\eqref{eq:H-IRLM}, with $\psi'=\psi$ and
\be
\label{eq:tp_U}
    U = \frac{4t^2}{\lambda}, \ \ t'=-\frac{4tw}{\lambda}.
\ee
The two resonant states are $\ket{101}$ ($\tilde{d}$ occupied) and $\ket{010}$ ($\tilde{d}$ empty). As flipping between these two states involves two tunneling amplitudes, $t' \propto t w$ with the energy denominator $\lambda$. The $U$ term is nothing but a scattering phase shift on the $\psi$ electrons which depends on the occupancy of the $\tilde{d}$-level.
Namely, according to $H_\text{IRLM}$ in Eq.~\eqref{eq:H-IRLM}, the two states of the resonant level yield a potential $\pm U/2$ on the lead fermions, which results in a phase shift $\pm \delta$ as in Eq.~\eqref{eq:delta-IRLM}. Indeed, substituting $\frac{\pi\nu U}{2}=\frac{\Gamma}{\lambda/2}$ from Eq.~\eqref{eq:tp_U} into Eq.~\eqref{eq:delta-IRLM} yields Eq.~\eqref{eq:delta}, i.e., $\delta=\arctan\frac{\Gamma}{\lambda/2}$. Note that the phase shift, like $U$ in Eq.~\eqref{eq:tp_U}, is calculated to zeroth order in $w$, and to leading order in $t$. Interestingly, however, we get the correct phase shift to all orders in $t$ (and to zeroth order in $w$), i.e., for the full range of $\lambda/\Gamma$.


For future reference, away from the particle-hole symmetric point, but with $|\epsilon_1|,|\epsilon_d|<\lambda/2$ so that the two ground states are still $\ket{101}$ and $\ket{010}$, we have to leading order
\begin{subequations}
\bea \label{eq:U_ed}
    U &=& +\frac{t^2}{\lambda/2+\epsilon_d}+\frac{t^2}{\lambda/2-\epsilon_d},\\
    t' &=& -\frac{tw}{\lambda/2+\epsilon_d}-\frac{tw}{\lambda/2-\epsilon_d}, \label{eq:tp}\\
    \epsilon_0 &=& \epsilon_1-\epsilon_d. \label{eq:eps0}
\eea
\end{subequations}
Note that in this case, according to Eqs.~\eqref{eq:delta} and \eqref{eq:delta-DD} we expect the phase shift to be
\be\label{eq:delta-epsd}
    \delta = \frac{1}{2}\arctan\frac{\Gamma}{\lambda/2+\epsilon_d}
           + \frac{1}{2}\arctan\frac{\Gamma}{\lambda/2-\epsilon_d}.
\ee
Here, in contrast to the $\epsilon_d=0$ case, substituting $U$ as in Eq.~\eqref{eq:U_ed} into Eq.~\eqref{eq:delta-IRLM} will only yield the expected $\delta$ to leading order in $t$. 


\subsection{Mapping to the IRLM: $\lambda \ll \Gamma$}\label{se:SW2}\label{se:1_2}
In the opposite limit $\lambda \ll \Gamma$ our reference state is a RLM decoupled from the double dot, i.e., the first line of Eq.~\eqref{eq:H}. Explicitly
\be
    H_\text{RLM} = H_\text{lead}[\psi] 
       + t[\psi^\dagger(0)d+\mathrm{H.c.}]
       .
\ee
Let us consider low energies compared to $\Gamma$. Then, the $d$-level is absorbed into the lead, changing its boundary condition. One can show that in this limit
\be
    H_\text{RLM} \overset{{T \ll \Gamma}}{\longrightarrow}  H_\text{lead}[\psi']
\ee
where $\psi'(x) = \psi(x)$ for $x<0$ and $\psi'(x) = -\psi(x) 
$ for $x>0$. Namely, the resonant level $d$ disappears from the model, as in the case of a quantum point contact detector, and we have an effective lead $H_\text{lead}[\psi']$. Thus, the charge of the $d$-level is not well defined, and the charge-stability diagram is irrelevant. The two level system in this limit corresponds to the two states of the double dot, which can be denoted as $\ket{n_1n_2}=\ket{01},\ket{10}$. Let us focus on the case $\epsilon_d=0$, with more details of the derivation provided in Appendix~\ref{se:appendix1}. One can show in this limit that the low energy operator identity holds~\cite{Sela_2009,sela2009nonequilibrium}
\be\label{eq:d}
    d \sim - \frac{2 \hbar v_F}{t} \psi'(x=0).
\ee
Introducing $\lambda$ perturbatively yields
\bea
    H &\to& H_\text{lead}[\psi'] + H_w \notag\\
       &+& 2\hbar v_F\frac{\lambda}{\Gamma} 
            {\psi'}^\dagger (0)\psi'(0)
            \big(d^\dagger_1 d_1 - \tfrac{1}{2}\big) .
\eea
Bosonizing this Hamiltonian with $\psi'(x)=F\frac{1}{\sqrt{2\pi a}}e^{i \phi(x)}$, where $ a=\hbar v_F/\Gamma$  is a short distance cutoff of this theory, and $F$ is a Klein factor, we obtain the spin-boson model precisely as in Ref.~\cite{Ma_2023} 
\be
    H \to \hbar v_F \int \frac{dx}{4\pi} (\partial_x \phi)^2  - \frac{\hbar v_F}{\pi}\overbrace{\frac{\lambda/2}{\Gamma}}^{\delta_\text{DD}} \partial_x \phi \sigma^z + w \sigma^x.
\ee
Then, we can refermionize to get the IRLM \cite{guinea1985bosonization,PhysRevB.93.165130}. One option is to perform a unitary transformation $H \to \hat{\mathcal{U}} H \hat{\mathcal{U}}^\dagger$ with $\hat{\mathcal{U}}=e^{i \sigma_z \delta_\text{DD}\phi(0)/\pi}$. This transformation completely removes the term proportional to $\lambda$ from the Hamiltonian and results in $\sigma^+ \to \sigma^+ e^{i 2 \delta_\text{DD} \phi(0)/\pi}$. Thus, the scaling dimension of the $w$ operator is $2(\delta_\text{DD} / \pi)^2$. However, in order to obtain the IRLM, one can instead demand that the scaling dimension of $w$ becomes that of free fermions, $1/2$. This is achieved with $\hat{\mathcal{U}}=e^{i \sigma_z \tilde{\delta} \phi(0)/\pi}$ with $\tilde{\delta}=\pi/2$, namely $\hat{\mathcal{U}}=e^{i \sigma_z \phi(0)/2}$. It results in the IRLM for the $\psi'$ fermions as in Eq.~\eqref{eq:H-IRLM} with
\begin{subequations}
\bea
    \delta &=& \frac{\pi}{2}-\frac{\lambda}{2\Gamma}, \label{eq:delta-LllG} \\
    t' &=& w \sqrt{2\pi a}=\sqrt{\pi}\frac{t w}{\Gamma}.\label{eq:tp-LllG}
\eea
\end{subequations}
Observe that Eq.~\eqref{eq:delta-LllG} is indeed the leading order expansion of Eq.~\eqref{eq:delta} in the limit of $\lambda\ll\Gamma$.

\begin{figure*}
\centering
\includegraphics[width=1\textwidth]{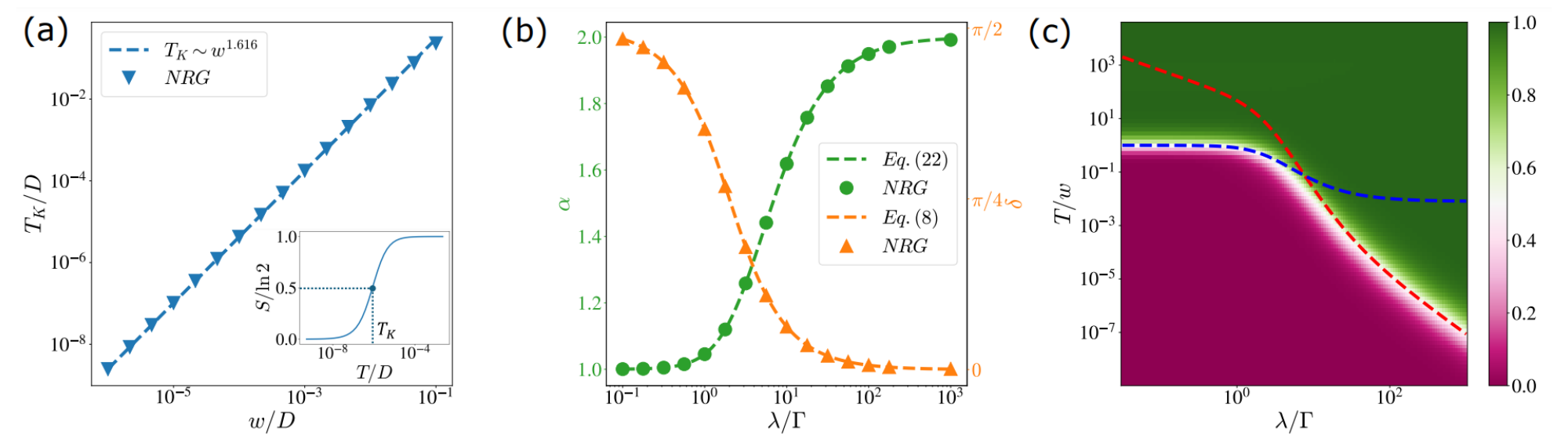}
    \vspace{-2\baselineskip}\phantomsubfloat{\label{fg:TK-w}}\phantomsubfloat{\label{fg:alpha-LoG}}\phantomsubfloat{\label{fg:S-T-LoG}}
    \caption{
    (a) $T_K$ as a function of $w$ for $\Gamma=0.01D$ and $\lambda=0.1D$ in units of the half-bandwidth $D$. 
    Here, $T_K$ is defined as the temperature at which the impurity entropy is halfway between $0$ and $\ln 2$, i.e., $S(T_K)=\frac{1}{2}\ln{2}$ (see inset for $w=10^{-4}D$). Fitting a power-law dependence $T_K \propto w^\alpha$, we obtain $\alpha=1.618$. 
    (b) Fixing $\Gamma=10^{-4}D$  and varying $\lambda$, we extract $\alpha$ as in (a), and invert Eq.~\eqref{eq:powerlaw} to get $\delta$, finding excellent agreement with Eqs.~\eqref{eq:alpha} and \eqref{eq:delta}, respectively.
    (c) Entropy for $w=10^{-4}D$, $\Gamma=0.01D$ and varying $\lambda$, as a function of the ratio $\lambda/\Gamma$ and temperature $T$ [inset to (a) corresponds to a vertical slice at $\lambda/\Gamma=0.1$]. Green corresponds to $S\approx\ln 2$, purple to $S\approx0$, and white to $S\approx\frac{1}{2}\ln 2$, tracking the evolution of $T_K$. The analytic expressions for $T_K$ are indicated by dashed lines in the two limits $\lambda\ll\Gamma$ (blue) and $\lambda\gg\Gamma$ (red).}
\label{fg:2}
\end{figure*}

\section{$T_K$ and the Scaling Exponent}\label{se:Tk}
From the previous two limits, we see that the effective low-energy model of a double dot coupled to a QD detector is the IRLM. However, depending on the ratio $\lambda/\Gamma$, the realization of the two states of the resonant level continuously changes: while for large $\lambda$ they correspond to the states denoted above as $\ket{101}$ and $\ket{010}$, as $\lambda \to 0$ they correspond to the two states of the double dot $\ket{01}$ and $\ket{10}$.  Generally, the tunneling term that flips between the states is $t' \propto w$, but the exact expression for $t'$ changes between these two limits. On the other hand, the expression for the phase shift $\delta$  in Eq.~\eqref{eq:delta} holds in both limits, and as we will now show numerically, for the full range of $\lambda/\Gamma$.

The IRLM gives rise to an emergent energy scale $T_K$ (the analog to the Kondo temperature in the Kondo model).
\be\label{eq:powerlaw}
    T_K\sim t'^{\alpha},\ \ \alpha =\frac{1}{1-\frac{1}{2}\big(1-\frac{2}{\pi}\delta\big)^2}.
\ee
The exponent $\alpha=1/(1-d)$ is related to the scaling dimension $d$ of the tunneling operator. The latter can be derived using standard RG analysis, yielding \cite{Camacho_2019} $d=(1-g)^2/2$ with $g=\frac{2}{\pi}\delta$. Substituting $\delta$ for $\epsilon_d=0$ from Eq.~\eqref{eq:delta} and using $\arctan x=\frac{\pi}{2}-\arctan\frac{1}{x}$, we obtain
\be\label{eq:alpha}
    \alpha_{\epsilon_d=0} =\frac{1}{1-\frac{1}{2}\Big(\frac{2}{\pi}\arctan\frac{\lambda/2}{\Gamma}\Big)^2}.
\ee

Consider the two limits of $\alpha$: At zero interaction $U=0$ (i.e., $\lambda \gg\Gamma$) we get $\alpha\to 2$ as  $T_k$ coincides with the bare width of the $\tilde{d}$-level, i.e., is quadratic in $t'$. In the opposite limit, of $U\to\infty$, the dot and the first site form a two-state system with splitting $t'$, and thus $\alpha\to 1$. In the original model, we have a decoupled double dot with Hamiltonian $H_w=w d_1^\dagger d_2+\mathrm{H.c.}$ and the energy scale $T_K \sim w$ is simply the bonding-antibonding splitting.

$T_K$ and $\alpha$ can be extracted from NRG simulations of the original double-dot model in Eq.~\eqref{eq:H} as follows. With NRG it is convenient to calculate the impurity entropy $S$ for any fixed  $\Gamma,\lambda,w$. Here we define $S$ as the difference between the thermodynamic entropy of the full system, and that of the detector, i.e., with an empty double dot. For large $T$ the two double-dot states are accessible and we have $S=\ln 2$, while below $T_K$ we observe their splitting and $S=0$. Thus, as shown in the inset of Fig.~\ref{fg:TK-w}, we identify $T_K$ as the midpoint $S(T_K)=\tfrac{1}{2}\ln 2$. For any fixed $\Gamma,\lambda$ we have $T_k\sim w^\alpha$ (as  $t'\propto w$), and so by varying $w$ and fitting we extract $\alpha$ [see Fig.~\ref{fg:TK-w}]. We then plot $\alpha$ and in Fig.~\ref{fg:alpha-LoG} as a function of the ratio $\lambda/\Gamma$ and find perfect agreement with Eq.~\eqref{eq:alpha}. For reference, we also plot $\delta$ by inverting the relation in Eq.~\eqref{eq:powerlaw} to demonstrate its agreement with Eq.~\eqref{eq:delta}.

While for $\alpha$ we had an analytical expression for the full parameter range, for $t'$ we only know the limit cases. In Fig.~\ref{fg:S-T-LoG} we plot the impurity entropy as a function of temperature and the ratio $\lambda/\Gamma$, with the color-scale indicating the value of $S$, and white corresponding to $T_K$. Following Ref.~\cite{Camacho_2019}, we explicitly rewrite Eq.~\eqref{eq:powerlaw} as
\be
    \nu'T_K = C(\alpha)(\nu't')^\alpha
    \label{eq:tk_analytic}
\ee
where $\nu'$ is the IRLM density of states, i.e., of $\psi'(0)$. The prefactor $C(\alpha)$ is of order one with a weak dependence on $\alpha$. Its value is chosen in the two limits as discussed in Appendix \ref{se:TK_const}.
The above equation is given in Ref.~\cite{Camacho_2019} for a lattice model, so that $\nu'$ and $t'$ must be redefined accordingly. This proves convenient for comparison with NRG, as the latter is also defined on a lattice. 

In the $\lambda\gg\Gamma$ limit $\psi'$ coincides with the bare $\psi$ so that $\nu'=\nu=1/\pi D$ where $D$ is the NRG high-energy cutoff. As the Schrieffer-Wolff analysis in Sec.~\ref{se:SW1} would look identical on a lattice, we take $t'$ as in Eq.~\eqref{eq:tp_U} where $t$ is now understood as the lattice version of the tunneling. This $T_K$ is marked in Fig.~\ref{fg:S-T-LoG} by a dashed red line, and displays excellent agreement down to $\lambda\approx\Gamma$. In the opposite limit of $\lambda\ll\Gamma$, the effective bandwidth is $\Gamma$ and so we take $\nu'=1/\pi\Gamma$. Transforming Eq.~\eqref{eq:tp-LllG} to a lattice model, we have  $t'\to t'/\sqrt{a}=\sqrt{2\pi}w$, where $a$ is the short distance cutoff of this theory. $T_K$ in this limit is marked by a dashed blue line. As $\alpha\approx 1$, and for an appropriate choice of the prefactor, we indeed observe $T_K\approx w$.

Observe that as we keep increasing the ratio $\lambda/\Gamma$, the energy scale $T_K$ keeps decreasing, although the limit $\alpha\to 2$ has already been saturated. This is because $t'$ keeps decreasing. It should not be confused with the localization transition, at which $T_K\to 0$ due to the divergence of $\alpha$ (at $\delta=\frac{\pi}{2}-\frac{\delta}{\sqrt{2}}$). For the latter, we expect $S=\ln 2$ at all temperatures above some critical $\lambda_c$.

\begin{figure*}
\centering
\includegraphics[width=\textwidth]{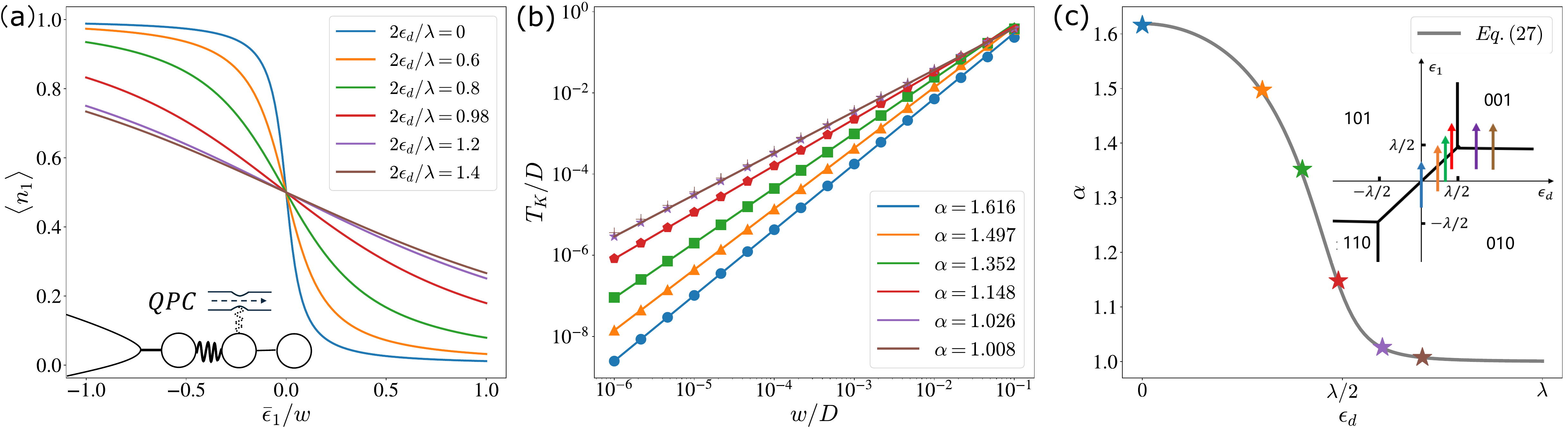}
    \vspace{-2\baselineskip}\phantomsubfloat{\label{fg:charge-eps1}}\phantomsubfloat{\label{fg:charge-fit}}\phantomsubfloat{\label{fg:charge-alpha}}
  \caption{
  (a)  Charging curve of QD1 at $T=0$, i.e., $\braket{n_1}$ as a function of $\epsilon_1$ for various $\epsilon_d$ along vertical cuts of the charge-stability diagram -- see inset to panel (c). The $x$-axis $\bar\epsilon_1$ is equal to $\epsilon_1$ up to an $\epsilon_d$-dependent shift such that $\braket{n_1}=0$ at $\bar\epsilon_1=0$. The charging curve can be measured by a charge detector that is weakly coupled to QD1 as shown in the inset. (b) Varying $w$ for fixed $\Gamma=0.01D$ and $\lambda=0.1D$ we fit $T_K\sim w^\alpha$ and extract $\alpha$. (c) The extract $\alpha$ for each $\epsilon_d$ is in perfect agreement with Eq.~\eqref{eq:alpha-epsd}.}
    \label{fg:3}
\end{figure*}

\section{Width of charging curve}
\label{se:charge-susceptibility}

We now discuss the charging curve of dot 1, i.e., $\braket{n_1}$ versus the detuning $\epsilon_1$. The width of the charge curve is a convenient way to study the localization phase transition, at which the charge curve becomes discontinuous with vanishing width~\cite{li2005hidden}. As discussed, in our model, the transition does not take place. Yet, as we will see we can study at what parameters regime the width becomes minimal, and we will also directly relate it to $T_K$.

Let us briefly discuss how to measure the charging curve. In the strongly interacting regime $\lambda\gg\Gamma$, the QD detector is part of the system, and so cannot be used to measure $\braket{n_1}$. Instead, we envision introducing a second charge detector that is only weakly coupled to QD1, as depicted in the inset to Fig.~\ref{fg:charge-eps1}. Then, by measuring the conductance through this second detector we obtain $\braket{n_1}$ with negligible effect on the system.

To gain an intuition for the charging curve, first consider the weakly interacting regime. For $\lambda\to0$ the double dot decouples, and for $T\ll w$ we have
\be
    \braket{n_1} = \frac{1}{2}-\frac{1}{2} \frac{\epsilon_1 }{\sqrt{4w^2+\epsilon_1^2}}
    .
\ee
Thus, the charge susceptibility, i.e., 
the slope of the charging curve at $\epsilon_1=0$ (for which $\braket{n_1}=1/2$), is
\be\label{eq:susc-n1}
    \left.-\frac{dn_1}{d\epsilon_1}\right|_{\epsilon_1\to 0}=\frac{1}{4w}.
\ee
In other words, the width of the charging curve is proportional to $w$. Any deviation from this behavior will be due to interaction with the detector. 

In order to consider larger $\lambda$, we turn to the IRLM. At low temperatures $T\ll T_K$, the charge susceptibly of the $\tilde{d}$-level is given by\footnote{The exact definition of $T_K$ can differ by factors of order one. Often the susceptibility is equated with $1/\pi T_K$ (see e.g., Ref.~\cite{Camacho_2019}), but here we chose $1/4T_K$ as in Ref.~\cite{PhysRevB.93.165130} to make contact with Eq.~\eqref{eq:susc-n1}. This definition of $T_K$ also differs by a multiplicative constant from the entropy half-height definition employed in the previous section.}
\be\label{eq:susc-ndt}
    \left.-\frac{dn_{\tilde{d}}}{d\epsilon_0}\right|_{\epsilon_0\to 0} = \frac{1}{4 T_K}.
\ee
As can be seen from the limit cases [see e.g., Eq.~\eqref{eq:eps0}], $\epsilon_0$ is equal to $\epsilon_1$ up to an $\epsilon_d$-dependent shift, which we find numerically by imposing $\braket{n_{\tilde{d}}}=1/2$ at $\epsilon_0=0$. We also note that in both limits $\braket{n_{\tilde{d}}}=\braket{n_1}$. Thus, up to a redefinition of $\epsilon_1\to\bar\epsilon_1$, such that $\braket{n_1}=1/2$ at $\bar\epsilon_1=0$, the measured charge susceptibility of Eq.~\eqref{eq:susc-n1} is equal to that of the IRLM in Eq.~\eqref{eq:susc-ndt}. In other words, the width of the charging curve is proportional to $T_K$. With $T_K\sim t'^\alpha$ and $t'\propto w$, we can use the $w$ dependence of the susceptibility to extract $\alpha$. As in the previous section, we can vary $\alpha$ according to Eq.~\eqref{eq:alpha} by varying the ratio $\lambda/\Gamma$. Here we take a different path, with similar results.

In Fig.~\ref{fg:charge-eps1} we plot the charging curve for vertical cuts of the charge-stability diagram, which cross the diagonal separating $\ket{010}$ from $\ket{101}$ or the horizontal line separating $\ket{010}$ from $\ket{001}$ [see inset to Fig.~\ref{fg:charge-alpha}]. Observe the narrowing of the width $\sim T_K$ as we approach the origin of the charge-stability diagram. Along the horizontal line, on the other hand, the width approaches $w$, as the QD detector is off-resonance. Focusing on the diagonal, we would like to use the narrowing to gauge the enhancement of $\alpha$, or the suppression of $\delta$ according to Eq.~\eqref{eq:delta-epsd} as $\epsilon_d\to 0$. However, as $T_K\sim t'^{\alpha}$, we need to also account for the trivial narrowing due to the suppression of $t'$ according to Eq.~\eqref{eq:tp} as $\epsilon_d\to 0$. To separate these effects, we can proceed as in Sec.~\ref{se:Tk}: we vary $w$ while fixing all other parameters, and then fit $\alpha$, as shown in Fig.~\ref{fg:charge-fit}. In Fig.~\ref{fg:charge-alpha} we observe excellent agreement with $\alpha$ as obtained by substituting $\delta$ from Eq.~\eqref{eq:delta-epsd} into Eq.~\eqref{eq:powerlaw}, or explicitly
\be\label{eq:alpha-epsd}
    \alpha  = \frac{1}{1-\frac{1}{2}\Big(
         \frac{1}{\pi}\arctan\frac{\epsilon_d+\lambda/2}{\Gamma}
        -\frac{1}{\pi}\arctan\frac{\epsilon_d-\lambda/2}{\Gamma}
    \Big)^2}.
\ee

Although we expect that $w$ can be varied in an experiment, its value must be measured (up to some multiplicative constant) to set the $x$ axis of Fig.~\ref{fg:charge-fit}. This can be achieved by assuming $\alpha=1$ for $\epsilon_d>\lambda/2$, and using a sweep of $w$ in that regime as the $x$ axis. Similarly, we can fix $\epsilon_d=0$ and vary $\lambda/\Gamma$ to change $\alpha$, while setting the $x$ axis as the width at the $\lambda\to 0$ limit. The requirement for a reference measurement of $w$ comes with the risk that model parameters might change between the two measurements. For example, as we vary $\epsilon_d$, we also change $\bar\epsilon_1$, which in turn could lead to a dependence of the bare $w$ on $\epsilon_d$.

A second major drawback of the above approach for extracting $\alpha$, as with any approach that relies on measuring $T_K$, is the dependence on $t'$ (and not the bare $w$). Typically, large $\alpha$ comes with a small $t'$, 
in which case the narrowing of the width could drop below the experimental accuracy due to the suppression of $t'$, regardless of $\alpha$. We point out that the narrowing due to the suppression of $t'$ makes it difficult to use such a probe to observe a localization transition (which in the model in this work cannot be reached). While at the transition $\alpha$ diverges and above it the charging curve is not continuous, it could be difficult to discriminate between such discontinuity and a trivial narrowing due to $t'$.

\section{Conductance}\label{se:spectral_function}
\begin{figure}
\centering
\includegraphics[width=.9\columnwidth]{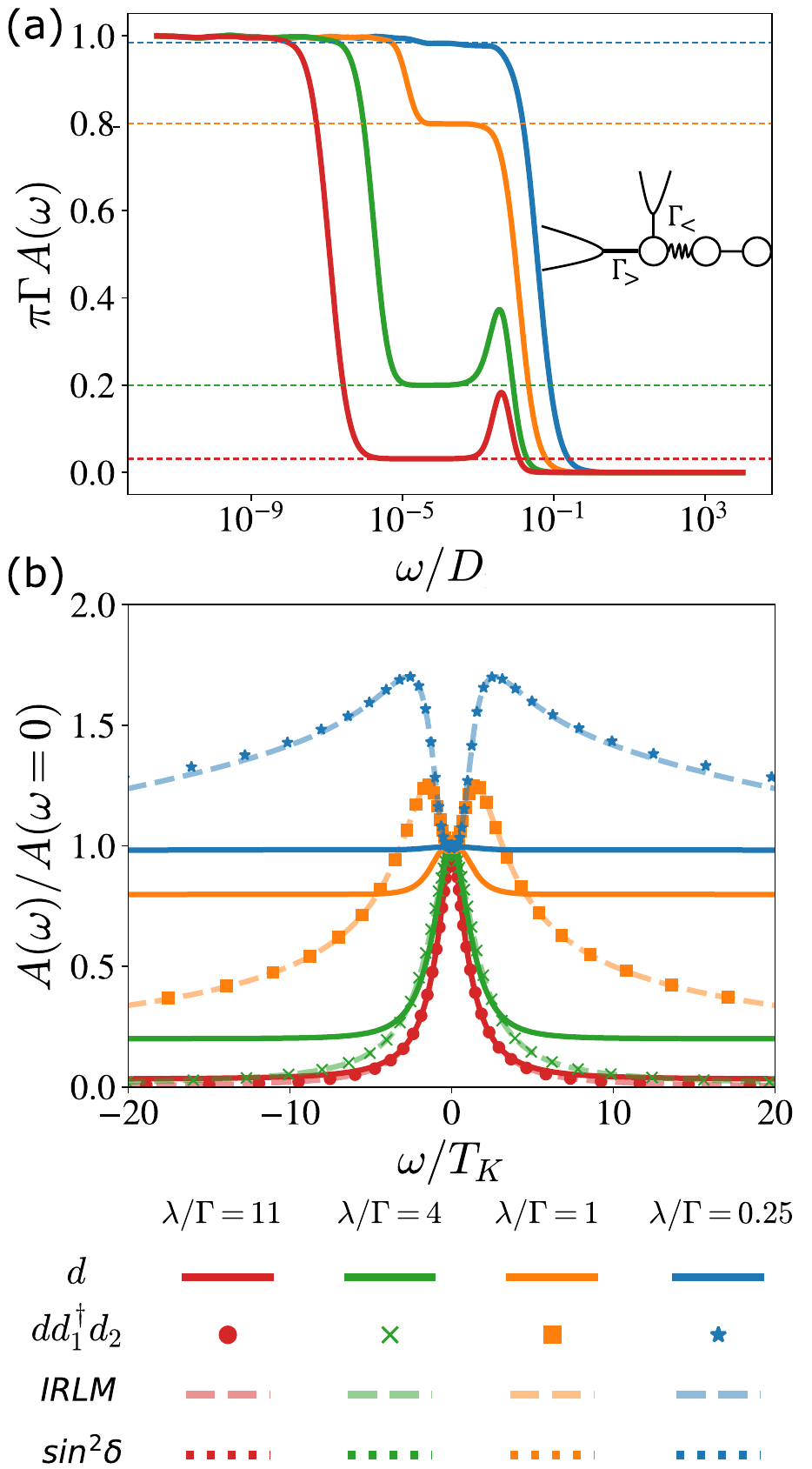}
\vspace{-2\baselineskip}\phantomsubfloat{\label{fg:spectral-d}}\phantomsubfloat{\label{fg:spectral-fit}}
  \caption{
  (a) The spectral function of the $d$-level (solid lines) for various $\Gamma$ while fixing $\lambda=0.01D$ and $w=10^{-5}D$. This spectral function can be probed by weakly coupling the detector QD to another lead as shown in the inset and measuring the conductance through it. Observe a plateau in the $T_K < \omega < \lambda/2$ regime with height $\sin^2\delta$ (dotted lines). (b) The spectral function of $d d^{\dagger}_1 d_2$ (markers) for the same parameters as in (a) and the spectral function of the $\tilde{d}$-level (dashed lines) in the IRLM, for the same values of $\delta$. The $x$-axis and $y$-axis are properly normalized such that the $\tilde{d}$-level spectral function depends solely on $\delta$. The $d$-level spectral function is also plotted (solid lines) for comparison.
  }
    \label{fg:4}
\end{figure}
We now turn to a different route. Instead of introducing a second detector, we suggest measuring the conductance through the original QD detector, which is tunnel-coupled to two leads as shown in the inset to Fig.~\ref{fg:4}. The couplings are taken to be very asymmetric $\Gamma_>\gg \Gamma_<$, with the potential of the strongly coupled lead fixed to the Fermi energy, and the full voltage bias $V$ falling on the weakly coupled lead. In this limit, the differential conductance $G(V)=dI/dV$ through the QD detector is given by \cite{meir1992landauer}
\be
    G(V)/G_0 =4\pi 
    \Gamma_<
    A(\omega=eV).
\ee
$G_0=e^2/h$ is the conductance quantum and $A(\omega)$ is the equilibrium spectral function (or local density of states) of the $d$-level
\be\label{eq:spectral}
    \!\!\!A(\omega)=-\frac{1}{\pi}\mathrm{Im}G_d^R
    \!(\omega)
    =\mathrm{Im}\frac{i}{\pi}\!\int_0^\infty\!\!\!\!\! dt e^{i\omega t}\!\braket{\{d(t),d^\dagger(0)\}}\!,\!\!\!\!
\ee
By construction, $A(\omega)$ depends only on the strongly-coupled lead so that $\Gamma=\Gamma_>+\Gamma_<\approx\Gamma_>$ and the QD potential $\epsilon_d$ is unaffected by the bias. 



Let us study the properties of $A(\omega)$, as plotted in Fig.~\ref{fg:spectral-d} for a single lead. We focus on the particle-hole symmetric point (the origin of the charge-stability diagram), for which $A(\omega)=A(-\omega)$. Fixing the interaction $\lambda$ and then double-dot tunneling $w$, we vary $\Gamma$ and rescale $A(\omega)$ accordingly. Observe three distinct features: a zero-bias peak, satellite peaks at $\pm\lambda/2$, and a plateau between the peaks.

The height of the zero-bias peak is fixed to $1/\pi\Gamma$, corresponding to perfect conductance. This behavior is clear in the $\lambda\to 0$ limit, where the QD detector is described by a resonant level and $A(\omega)=\frac{\Gamma/\pi}{\omega^2+\Gamma^2}$. In the IRLM picture, this corresponds to the strong-coupling limit ($U\to\infty$) with $\delta\to\pi/2$, for which the zero-bias conductance is given by $G=G_0\sin^2\delta$. Importantly, at the particle-hole symmetric point, the IRLM with any $U>0$ flows to a strong-coupling fixed point, i.e.,  a Fermi-liquid with $\delta=\pi/2$. Thus, the height of the zero-bias peak is independent of the bare $\delta$ and is equal to $1/\pi\Gamma$ for any $\lambda/\Gamma$ and $w$. As this holds only below $T_K$, the width of the peak does depend on the model parameters and is given by $\sim\min\{\Gamma,T_K\}$.

Next, consider the plateau in the $T_K<\omega<\lambda/2$ regime. Its height is given by
\be\label{eq:G-plateau}
    G = \braket{n_1}G_{\epsilon_d+\frac{\lambda}{2}}
      +(1-\braket{n_1})G_{\epsilon_d-\frac{\lambda}{2}}
      \overset{\!\!(\epsilon_d=0)\!\!}{=} G_0\sin^2 \delta,
\ee
with $\delta$ the bare phase shift according to Eq.~\eqref{eq:delta}. We understand this conductance to be generated by a probabilistic mixture of an electron present or absent at QD1, thereby applying a potential $\epsilon_d\pm\lambda/2$  to the $d$-level and yielding a conductance 
\be
    G_{\epsilon_d\pm\lambda/2} = G_0\sin^2[\delta(\epsilon_d\pm\lambda/2,\Gamma)-\pi/2],
\ee
according to Landauer's formula. At $\epsilon_d=0$ we have $\braket{n_1}=1/2$ for any temperature by particle-hole symmetry, yielding the right-hand side of Eq.~\eqref{eq:G-plateau}. Thus, by measuring this plateau and inverting Eq.~\eqref{eq:G-plateau} we get $\delta$ (and $\alpha)$ without varying $w$. 
We point out that the occupation $\braket{n_1}$ in Eq.~\eqref{eq:G-plateau} is taken at $T=\omega>T_K$. Although $\braket{n_1}=1/2$ for $T<T_K$ at any point along the diagonal, above $T_K$ it is constant only at the particle-hole symmetric point. Thus, for any $\epsilon_d\neq 0$ we will not observe plateaus due to the temperature dependence of $\braket{n_1}$.

Let us now compare with the IRLM $\tilde d$-level spectral function. The latter is discussed in detail in Ref.~~\cite{Camacho_2022}. For $\lambda\gg\Gamma$ the Schrieffer-Wolff mapping yields $\tilde d = d d_1^\dagger d_2$. Thus, in Fig.~\ref{fg:spectral-fit}
we plot the spectral function obtained by substituting $d\to d d_1^\dagger d_2$ into Eq.~\eqref{eq:spectral}. Following Ref.~\cite{Camacho_2022}, we rescale the $x$-axis by $T_K$ and normalize the spectral function such that $A(\omega=0)=1$. Under such a rescaling we observe excellent agreement with the universal IRLM curve for the spectral function of $\tilde{d}$, that depends only on $\delta$ (dashed lines) \cite{Camacho_2022}. Interestingly, this agreement holds even for $\Gamma\gtrsim \lambda$, which is presumably beyond the validity range of the Schrieffer-Wolff mapping. In the extreme limit $\lambda/\Gamma\to\infty$ (or $U=0$) we expect $d\sim\tilde d$. Indeed for the smallest $\Gamma$ (in red), we observe that the spectral functions agree up to normalization, i.e., correspond to a Lorentzian of width $T_K\sim t'^2$. However, for any finite $\lambda/\Gamma$ the two spectral functions deviate from each other. Thus, although the IRLM spectral function does emerge in the considered double-dot system, we leave the question of how to measure it for future work.

\section{Summary}
\label{se:summary}
We considered a double quantum dot that is monitored by a charge detector. In a recent paper \cite{Ma_2023} we showed that upon increasing the electrostatic coupling to the detector, this model displays a localization transition. However, experimentally it is not clear how to estimate this electrostatic coupling. Here, we focused on the case where the double dot interacts with a spinless QD detector. Our main finding is that the resulting model maps to the well-known interacting resonant level model (IRLM). 

We then applied results on the IRLM for the double-dot system. Particularly the IRLM has a dynamic energy scale, $T_K$, at which the entropy drops to zero, which also determines the charge susceptibility of the QD system, and the conductance of the QD detector. All of these quantities are experimentally accessible allowing to measure $T_K$. Importantly, $T_K$ scales as a power law of the tunneling, and this power law allows to extract directly the coupling between the detector and the system, even when the system is far from the localization transition. 

The IRLM with repulsive interaction $U>0$ does not display the localization transition even upon increasing $U$ to infinity. In our previous work~\cite{Ma_2023} we obtained a localization transition by enhancing the electrostatic coupling beyond the first site of the detector while here we realistically assumed that the double dot interacts only with the QD of the detector. An experimentally plausible way to enhance the interaction and observe the transition is to add multiple detectors, which also requires to include spin in the analysis. However, in that case, more complex effects such as the Kondo effect in the detector take place. The prospect of observing the transition in realistic systems is left for future work~\cite{inprep2}.

\begin{acknowledgments}
We would like to thank Josh Folk for insightful discussions and for providing us with the motivation
for this work. We gratefully acknowledge support from the European Research Council (ERC) under the European Union Horizon 2020 research and innovation program under grant agreement No.~951541. 
\end{acknowledgments}

\begin{appendix}
\section{Derivation of Eq.~\eqref{eq:d}}
\label{se:appendix1}
This appendix provides supplementary information to Sec.~\ref{se:1_2}, where we derived the effective model in the small interaction limit $\lambda \to 0$. Treating $\lambda$ perturbatively, we start from $\lambda=0$ for which we have the RLM
\be
H_\text{RLM} = -i\hbar v_F \int dx  \psi^\dagger \partial_x \psi + t[\psi^\dagger(0) d+\mathrm{H.c}.]+\epsilon_d d^\dagger d.
\ee
It can be diagonalized into $H=\sum_k \epsilon_k \psi^\dagger_k \psi_k$ with eigenmodes $\{\psi_{k} ,\psi^\dagger_{k'} \}=\delta_{kk'}$. We can express the field $\psi(x)$ and the $d$ operator as a mode expansion
\be
\psi(x)=\sum \varphi_k(x) \psi_k, \quad 
d=\sum_k f_k \psi_k.
\ee
Expressing the Heisenberg equations of motion for the operators $\psi(x)$ and $d$ in terms of the mode operators $\psi_k$, we obtain
the pair of equations
\bea
-i\hbar v_F \partial_x \varphi_k(x)+ t \delta(x) f_k &=&\epsilon_k \varphi_k(x),  \nonumber \\
\epsilon_d f_k + t \varphi_k(0) &=&\epsilon_k f_k. 
\eea
The solutions with $\epsilon_k=\hbar v_F k$ have the form $\varphi_k(x)=e^{-i \delta_k}e^{ikx}$ for $x<0$ and  $\varphi_k(x)=e^{+i \delta_k}e^{ikx}$ for $x>0$. Then $\varphi_k(0)=\cos \delta_k$ and $-i \int_{-\epsilon}^\epsilon dx \partial_x \varphi_k(x) \to_{\epsilon \to 0} 2 \sin \delta_k$.
We obtain
\be
\tan \delta_k = -\frac{t^2}{2 \hbar v_F (\epsilon_k-\epsilon_d)}.
\ee
For $k \approx 0$ we can write more conveniently $\delta_k=\frac{\pi}{2}+\tilde{\delta}_k$ with 
\be
\tan \tilde{\delta}_k = \frac{  2 \hbar v_F (\epsilon_k-\epsilon_d)   }{t^2}.
\ee

We now define an effective lead fermion $\psi'(x)=\sum_k \varphi'_k(x) \psi_k$ where 
$\varphi_k'(x)=e^{ikx}$. 
Next we can express the $d$ operator using the equations of motion
\be
d=\sum_k f_k \psi_k= \sum_k \frac{t \varphi_k(0)}{\epsilon_k-\epsilon_d}.
\ee
Using $ \varphi_k(0) = \cos \delta_k \sim - \sin \tilde{\delta} \sim 
\frac{2 \hbar v_F}{t^2} (\epsilon_k-\epsilon_d)$, we obtain
\be
d=-\psi'(0) \frac{2 \hbar v_F}{t}
.
\ee
We thus obtain
\be
H=
\lambda {:}{\psi'}^\dagger \psi' {:}\frac{2\hbar v_F}
{\Gamma} (d^\dagger_1 d_1-1/2), 
\ee
where the operator $(d^\dagger d-1/2)$ became the normal-ordered density operator ${:}{\psi'}^\dagger \psi'{:}$. We can write this using bosonization as
\be
H=\hbar v_F \int \frac{dx}{4\pi} (\partial_x \phi)^2 + 2 \hbar v_F\frac{\lambda }{\Gamma} \frac{\partial_x \phi}{2\pi} 
\frac{\sigma^z}{2} + w \sigma^x.
\ee
This maps to the spin-boson model (using the same notation of $\delta_\text{DD}$ as in Ref.~\cite{Ma_2023}) with coupling corresponding to 
\be
\delta_\text{DD}=\frac{\lambda}{2\Gamma}. 
\ee
Clearly this is consistent with Eq.~(\ref{eq:delta}) in the main text, using
\be
\delta =   \arctan \frac{\Gamma}{\lambda/2}=\frac{\pi}{2}- \arctan \frac{\lambda/2}{\Gamma}.
\ee

\section{The constant in Eq.~\eqref{eq:tk_analytic}}\label{se:TK_const}
In the limit $\lambda/\Gamma=\infty$, our model can be mapped to the non-interacting resonant level model, which is exactly solvable. Specifically, the impurity entropy is given by
\be
S(T) = \ln{2} - \frac{1}{2\pi}\int_0^{\infty}dx \frac{x}{\cosh^2\frac{x}{2}}\arctan\frac{\Gamma}{Tx}.
\ee
Solving the Equation $S(T_K)=\frac{1}{2}\ln 2$
, we get $T_K\approx 0.508\Gamma$. In this limit, Eq.~(\ref{eq:tk_analytic}) predicts $T_K=C\pi\nu't'^2=C\Gamma'$, this fixes the constant to be $0.508$.

In the opposite limit $\lambda/\Gamma=0$, the double dot and the QD detector are decoupled, the entropy of a double dot is given by
\be
S(T) = \ln{2} + \ln{\cosh \frac{w}{T}} - \frac{w}{T}\tanh\frac{w}{T}
\ee
Again solving the Equation $S(T_K)=\frac{1}{2}\ln{2}$, we get $T_K\approx 1.045 w$. 
Eq.~(\ref{eq:tk_analytic}) predicts $T_K=C\sqrt{2\pi}w$ as $t'=\sqrt{2\pi} w$ in this limit, this fixes $C\approx 0.399$.
\end{appendix}
\bibliography{references1}

\clearpage
\end{document}